\documentclass[prla,a4paper,superscriptaddress,twocolumn,showpacs,amsmath,amssymb,floatfix,amstex]{revtex4}

\usepackage[dvips]{graphicx}
\usepackage{bm,pstricks,longtable}

\input{epsf}
\newcommand{\ket}[1]{| #1\rangle}                       %

\newcommand{\bra}[1]{\langle #1}                        %

\begin{document}

\title{Effects of imperfections for Shor's factorization algorithm}


\author{Ignacio Garc\'ia-Mata}
\affiliation{\mbox{Laboratoire de Physique Th\'eorique, UMR 5152 du CNRS, 
Universit\'e  Paul Sabatier, 31062 Toulouse Cedex 4, France}}
\author{Klaus M. Frahm} 
\affiliation{\mbox{Laboratoire de Physique Th\'eorique, UMR 5152 du CNRS, 
Universit\'e  Paul Sabatier, 31062 Toulouse Cedex 4, France}}
\author{Dima L. Shepelyansky}
\homepage[]{http://www.quantware.ups-tlse.fr}
\affiliation{\mbox{Laboratoire de Physique Th\'eorique, UMR 5152 du CNRS, 
Universit\'e  Paul Sabatier, 31062 Toulouse Cedex 4, France}}

\date{January 23, 2007}

\begin{abstract}
We study effects of imperfections induced
by residual couplings between qubits on the accuracy
of Shor's algorithm using numerical simulations
of realistic quantum computations with up to 30 qubits.
The factoring of numbers up to $N=943$
show that the width of peaks, which frequencies
allow to determine the factors, grow exponentially with
the number of qubits. However, the algorithm remains
operational up to a critical coupling strength $\epsilon_c$
which drops only polynomially with $\log_2 N$.
The numerical dependence of $\epsilon_c$ on $\log_2 N$
is explained by analytical estimates that
allows to obtain the scaling for functionality of Shor's
algorithm on realistic quantum computers
with a large number of qubits.
\end{abstract}


\maketitle
\section{Introduction}

It is quite clear that the main interest to the quantum computing 
has been generated by Shor's factorization algorithm \cite{shor1994}
which has exponential efficiency gain compared to any known classical
algorithm. Indeed, Shor's algorithm allows to find the factors
of a large number $N$ with $O(\ln^3 N)$ quantum gates
while all known classical algorithms require a number of operations that
grows exponentially with $\ln N$ (see e.g. review and relevant Refs.
in \cite{chuang}). Without any doubt this result 
has a fundamental importance
from a mathematical view point. However, its implementation in real
life requires an understanding of effects of imperfections and errors
unavoidably present in any real physical realization of the algorithm
on a realistic quantum computer. Again, here a mathematician
can be satisfied by a mathematical statement that 
in quantum computations the
errors grow not faster than quadratically with the number of performed
quantum gates (see \cite{chuang}) and thus the global accuracy of 
the algorithm is good enough if the norm of errors
in each quantum gate is sufficiently small. However, 
a physicist generally would like
to see more concrete and realistic estimates of the algorithm 
accuracy. Unfortunately, direct experimental verification
of the accuracy for a large number of gates 
and qubits is not possible at present.
Indeed, the most advanced quantum computation of Shor's algorithm
has been done on a 7-qubit NMR-based quantum computer that allowed 
to factorize only a rather small number $N=15$
(even if certain simplifications of the original algorithm
have been used) \cite{chuang1}. 

Therefore, the only possibility remaining is the method of numerical
simulations testing various types of realistic errors and imperfections.
The first steps in this direction have been done in
\cite{cirac1995,paz1996,paz1997}.
A number of interesting effects of errors on the accuracy
of Shor's algorithm has been found in these pioneering works
but the factorized number was still $N=15$ and therefore 
it was not possible to determine the accuracy scaling 
at large values of $N$. 
More recently, additional numerical studies 
have been performed to investigate the effects of
finite accuracy in quantum phase rotations 
of the quantum Fourier transform (QFT) algorithm
used in Shor's factorization \cite{china,hollenberg1},  
dynamical phase errors in Shor's algorithm
with $N$ up to 33 \cite{nori} and 
discrete qubit flip errors 
\cite{hollenberg2} with $N$ up to 247.
In the latter case the QFT part of Shor's algorithm
has been performed in a semiclassical way using the one qubit
control trick (see e.g. \cite{mosca,zalka,plenio,beauregard})
while the modular multiplication
has been been performed with up to 20 qubits including
the workspace using the circuit described in 
\cite{hollenberg3}.
 
In this work we perform extensive numerical simulations 
investigating effects of imperfections
on the accuracy of Shor's algorithm factorizing
numbers up to a maximal value $N=943$ using up to $L=30$ qubits.
We concentrate our studies on the case of static imperfections
which induce static one-qubit energy shifts and residual
static couplings between qubits following the lines started in
\cite{georgeot2000}. This type of imperfections is especially
important since generally the errors produced in this case are accumulated
coherently and lead to a more rapid drop of 
fidelity and accuracy of quantum
computations compared to the cases of noisy unitary errors
in quantum gates \cite{benenti2001,frahm2004}
and dissipative decoherence \cite{carlo,lee,zhirov}.
It is known that for the quantum algorithms
simulating problems of quantum chaos periodic in time
the Floquet eigenstates are exponentially sensitive
to static imperfections \cite{benenti2002}.
Due to that the study of their effects on the accuracy of 
Shor's algorithm becomes especially relevant since
recently it has been shown that certain blocks
of the Shor algorithm are characterized by the 
properties of quantum chaos \cite{indians}.
Also it is important to note that Shor's algorithm
is essentially based on a determination of 
a certain frequency of return. 
In some cases, like in the Grover algorithm,
such a frequency can be exponentially sensitive
to static couplings \cite{pomeransky}
that makes the investigation of static imperfections effects
in Shor's algorithm even more important.

Thus, in the present work we present the first numerical studies 
of how the static imperfections affect the accuracy of Shor's algorithm.
Our aim is to determine the parametric dependence
of the accuracy on the imperfection strength, number of qubits
and number of gates.  
For this we use a simplified but generic model
of imperfections which can be applied to various 
implementations of Shor's algorithm discussed in the literature 
\cite{paz1996,vedral,beckman,zalka,gossett,beauregard,draper,meter,zalka1}.

The paper has the following structure:
Section II gives a brief description of ideal
Shor's algorithm, Section III describes
the model of errors introduced by static imperfections,
the results of numerical studies are presented in Section IV 
and the discussion of the results
is given in Section V.

\section{Ideal realization of Shor's algorithm}
First we briefly describe the main structure of 
Shor's algorithm \cite{shor1994}  factorizing
a large integer number $N$ using parallelism
of many-body quantum evolution. Following Shor 
we choose a random number $x$ relatively prime to $N$  and calculate its 
{\em order} $r$ (also called {\em period}) defined as the minimal 
positive integer value such that
\begin{equation}
\label{eq:order}
x^r\equiv 1 \mod N.
\end{equation}
Once $r$ is known there is a high probability to obtain two non-trivial 
factors of $N$ by a classical computation in polynomial time 
(in the number of binary digits of $N$). This procedure 
fails in rare cases \cite{shor1994} 
and in such a case one has simply to chose 
a different value of $x$ and restart again.

The difficult task is to compute the order $r$ and this task 
can be efficiently achieved by Shor's algorithm provided we have a reliable
quantum computer with a sufficient number of qubits at our disposal. This 
algorithm requires an $L$-qubit state 
composed of two quantum registers which we will call the {\em control 
register} (with $n_l$ qubits) and the {\em computational register} (with 
$n_q=L-n_l$ qubits). We associate to the basis states of both registers 
integer numbers by:
\begin{equation}
\label{eq:reg_number}
\ket{\tilde l}=\ket{\alpha_{n-1}}_{n-1}\cdot\ldots\cdot \ket{\alpha_0}_0
\end{equation}
where in binary representation
\begin{equation}
\label{eq:reg_number2}
\tilde l =\alpha_0+2\alpha_1+\ldots+2^{n-1}\alpha_{n-1}
\end{equation}
and $n=n_l$ for the control register or $n=n_q$ for the computational 
register. Here $\ket{\alpha_j}_j$ represents the $j$th qubit of the register 
and $\alpha_j\in\{0,\,1\}$. 
In order to factorize a number $N$ one needs to choose 
$n_l$ and $n_q$ such that
$2^{n_q}> N$ and $Q\equiv 2^{n_l}>N^2$, 
therefore typically $n_l \approx 2 n_q$. 

We first prepare the initial state
\begin{equation}
\ket{\psi_0}=\ket{0}_{n_l}\ket{1}_{n_q}
\end{equation}
and then apply single qubit Hadamard gates to every qubit in the control 
register and get (dropping subscripts)
\begin{equation}
\ket{\psi_1}=\frac{1}{\sqrt{Q}}\sum_{a=0}^{Q-1}\ket{a}\ket{1} \; .
\end{equation}
The principal idea of Shor's algorithm is the observation that one can 
construct 
a combination of quantum gates, acting on both registers, that performs 
for all $a=0,\ldots,\,Q-1$ simultaneously the operation~:
\begin{equation}
\label{eq:operation}
\ket{a}\ket{1}\ \to\ \ket{a}\ket{x^a \mod N}
\end{equation}
which gives the state
\begin{equation}
\ket{\psi_2}=\frac{1}{\sqrt{Q}}\sum_a\ket{a}\ket{x^a \mod N} \; .
\end{equation}
Then, after obtaining the state $\ket{\psi_2}$,
we apply the QFT \cite{chuang}
to the control register 
\begin{equation}
\ket{\psi_3}=\frac{1}{Q}\sum_{c=0}^{Q-1} \sum_{a=0}^{Q-1} 
e^{i 2\pi ac/Q}\ket{c}\ket{x^a \mod N}
\end{equation}
and measure both arguments to get
\begin{eqnarray}
\nonumber
&&P(c,x^k) \equiv \left|\bra{\psi_3}\ket{c}\ket{x^k \mod N}\right|^2 \\
\label{eq:prob}
&&=\left|\frac{1}{Q}\sum_{\bar{a}:x^{\bar a}\equiv x^r} e^{i 2\pi\, c \bar{a}/Q}\right|^2
\end{eqnarray}
where $k=0,\ldots,\,r-1$ is arbitrary and the sum over $\bar{a}$ runs over 
all values such that $x^{\bar{a}}\equiv x^k \mod N$. Therefore 
$\bar{a}=r\nu+k$ where $\nu=0,\ldots,M_k-1$ and 
$M_k\equiv[(Q-k-1)/r]+1$ 
and the evaluation of the sum yields:
\begin{equation}
\label{eq:probresult}
P(c,x^k)=\frac{1}{Q^2}\frac{\sin^2(M_k\pi cr/Q)}{\sin^2(\pi cr/Q)} \; .
\end{equation}
This function only depends weakly on the choice of $k$ (since $Q>N^2$ and 
$N>r>k$ such that $Q\gg k$ and $M_k\approx Q/r\gg 1$ 
is nearly constant in $k$) and 
as a function of $c$ it has $r$ equidistant strongly localized 
peaks of width unity, of height $M_k^2/Q^2\approx 1/r^2$ and located at 
$m Q/r$ with $m=0,1,\ldots,r-1$. 

If the algorithm is run
by an {\em ideal\/} quantum computer, then with a very high probability the 
outcome of a measurement will be given by an integer value of $c$ which is very 
close to one of the peaks $mQ/r$. Thus, using a continuous fraction expansion 
we can determine the rational number $p/q$ closest to $c/Q$ with a denominator 
smaller than $N$. Here the choice $Q>N^2$ ensures that there is at most one 
such number inside the peak and therefore $p/q$ coincides with $m/r$. 
Furthermore the position number $m$ of the peak is quite random and if by 
chance $m$ is relatively prime to $r$ one obtains directly $r=q$ and the 
algorithm succeeds. However, if $m$ and $r$ have a common divisor larger than 
unity we have $q=r/\gcd(m,r)<r$ and the algorithm did not succeed. 
Therefore one has to check classically if the candidate ``$q$'' for $r$ is 
indeed a solution of $x^q=1 (\mod N)$. This fortunately can be done in a 
polynomial time. In case of failure the algorithm has to be repeated and even 
though the probability of success is not very high one obtains after a few 
(${\cal O}(\log\log r)$) measurements \cite{shor1994} the correct value $q=r$. 

Practically it is more convenient to measure only the control register which 
provides $c$ with the total probability:
\begin{equation}
\label{eq:probtot}
P(c)=\sum_k P(c,x^k)\approx r\,P(c,x^k) .
\end{equation}
We note that the dependence of $M_k$ on $k$ 
in Eq.~(\ref{eq:probresult}) 
is rather weak and therefore 
the above procedure to determine 
the minimal period $r$ remains the same.  

However, this description of the algorithm still lacks some precision how 
to implement the operation described in equation (\ref{eq:operation}). Suppose 
we are able to perform on the computational register the multiplication 
by $x \mod N$:
\begin{equation}
\label{eq:multiply}
\ket{y}\ \to\ U_{\rm mult}(x)\ket{y}\equiv \ket{(yx) \mod N}
\end{equation}
by some unitary operator. Of course this operator cannot be unitary if we 
require this for all values $y=0,\ldots 2^{n_q}-1$ 
simply because the classical 
application $y\to (xy)\mod N$ is not unique on this set (unless $N=2^{n_q}$ 
which is of no interest). If we require that 
$x$ and $N$ are relatively prime then this application is unique at least for 
$y=0,\ldots N-1$ and for $y=N,\ldots,2^{n_q}-1$ we have to complete it in some 
unique way, for example by: $y\to y$ if $y\ge N$. Therefore we define 
the quantum multiplication operator by $x\mod N$ by:
\begin{equation}
\label{eq:quantmultdef}
U_{\rm mult}(x)\ket{y}\equiv\left\{
\begin{array}{ll}
\ket{(yx) \mod N} &\ ,\ y=0,\ldots N-1 \\
\ket{y} &  \ ,\  y=N,\ldots,2^{n_q}-1\\
\end{array}\right.
\end{equation}
The states $\ket{y}$ with $y\ge N$ are in principle not relevant for the 
{\em ideal} Shor algorithm because they are never populated in the 
perfect computation and the effect 
of the quantum gates on these states is rarely discussed in the
literature \cite{shor1994,paz1996,vedral}. 
However, they are important to ensure overall unitarity 
and they may be very well  populated if the quantum computation is subjected 
to errors or imperfections. Furthermore, we note that in the definition 
(\ref{eq:quantmultdef}) we could in principle replace the unit-operator 
acting on the 
non-relevant states by an arbitrary unitary operator (acting on a space of 
dimension $2^{n_q}-N$) provided that we do not mix relevant ($y<N$) and 
non-relevant states ($y\ge N$).

We now introduce the controlled multiplication operator 
$U^{(j)}_{\rm Cmult}(x)$ 
acting on both registers (control and computational register) applying the 
simple multiplication (\ref{eq:quantmultdef}) on the computational register 
if and only if the $j-$th qubit of the control register is $\ket{1}$. 
Developing 
$a=\sum_{j=0}^{n_l-1} a_j\,2^j$ with $a_j\in\{0,\,1\}$ we see that the 
operation (\ref{eq:operation}) can be performed by the unitary operator:
\begin{equation}
\label{eq:fullmult}
U_{\rm Fmult}(x)=\prod_{j=0}^{n_l-1} U^{(j)}_{\rm Cmult}\left(x^{2^j} 
\mod N\right)
\end{equation}
since 
\begin{equation}
x^a =\prod_{j=0}^{n_l-1} \left(x^{2^j} \right)^{a_j}
=\prod_{j=0,a_j=1}^{n_l-1} x^{2^j} \ 
\end{equation}
and where in the last equation every multiplication is taken modulo $N$. 

\begin{figure}[h]
\epsfxsize=0.95\hsize
\epsffile{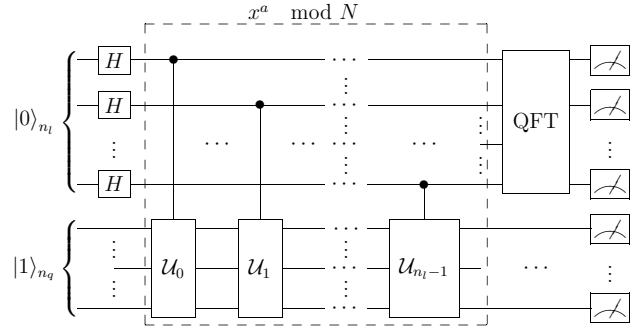}
\caption{\label{fig1}
Quantum circuit of Shor's algorithm on an ideal quantum computer with 
the quantum multiplication operator 
${\cal U}_j=U_{\rm mult}\big(x^{2^j} \mod N\big)$ as defined in 
(\ref{eq:quantmultdef})
}
\end{figure}
Fig. \ref{fig1} shows the schematic quantum circuit of Shor's algorithm 
on an ideal quantum computer in terms of the quantum multiplication operator 
(\ref{eq:quantmultdef}). To complete an explicit implementation one has 
to show that this operator can be 
realized in terms of elementary one or two qubit quantum-gates. We do not 
enter into details here and mention as examples the important works 
\cite{paz1996,vedral}  that 
provided explicit implementations of the quantum multiplication by 
$x \mod N$ using ${\cal O}(n_q^2)$ 
elementary gates. These implementations require also 
additional work space qubits which are initially $\ket{0\ldots 0}$ and must 
remain so after completion of this 
operator, i.e. the implementations must eventually provide code to 
reversibly ``erase'' 
the additional work space qubits. We assume that there are no
errors inside this additional work space and no errors
coupling it to the control and computational registers. In this way
we may restrict our consideration only to $L=n_q+n_l$ qubits.

\section{Shor's algorithm with static imperfections}
\label{sec3}

We now turn to Shor's algorithm in the case of static imperfections
\cite{georgeot2000} generated by residual 
couplings between  qubits and energy level shifts.
The effects of these imperfections 
and their numerical modeling have been considered
in detail in \cite{frahm2004} on examples of quantum chaos algorithms
(see also Refs. in \cite{frahm2004} on other works).
There it has been shown that effects of static residual couplings
can be modeled
by an additional unitary rotation acting between two arbitrary gates~: 
$U_s=e^{i \delta H}$. Here $\delta H$ represents the Hamiltonian 
due to the residual static couplings between qubits
which provides a non-trivial evolution 
of the state stored in the quantum register even in absence of any quantum 
gate. In this approach the quantum gates
are considered to be exactly  ideal.
In principle $\delta H$ may couple all qubits in the control register, 
in the computational register and in the additional work space 
necessary for the concrete implementations of the quantum multiplication 
operator (\ref{eq:quantmultdef}). However, in this work we use 
a simplified error model in which $\delta H$ couples only the qubits in the 
computational register and therefore in Shor's algorithm the initial 
Hadamard gates or the final quantum Fourier transform are not affected by 
these errors. In principle the quantum Fourier transform is considered as 
relatively stable with respect to errors \cite{cirac1995} and the number of
Hadamard gates $n_l$ is relatively small.

Furthermore we do not consider a specific implementation of the quantum 
multiplication operator (\ref{eq:quantmultdef}), we only assume that it can be 
written as a product
\begin{equation}
\label{multproduct}
U_{\rm mult}(x)=U_{n_m}\cdot \ldots\cdot U_2\cdot U_1
\end{equation}
where $U_j$, $j=1,\ldots, n_m$ are the elementary quantum gates which 
constitute this operator and $n_m={\cal O}(n_q^2)$ is the number of these 
elementary gates. A specific choice of $U_j$ depends on
the classical variable $x$, and also on $N$, and since the value of $x$
significantly affects the algorithm implementation
we have a different set of gates $U_j$ for each $x$ (and $N$).

Thus, in presence of static imperfections 
the quantum multiplication operator $\tilde U_{\rm mult}(x)$ has the form:
\begin{equation}
\label{multproduct_static}
\tilde U_{\rm mult}(x)=U_{n_m}\cdot e^{i\delta H}\cdot \ldots\cdot 
U_2\cdot e^{i\delta H}\cdot U_1\cdot e^{i\delta H}\ .
\end{equation}
We now introduce 
an effective perturbation operator for the full multiplication operator by: 
\def\X{{\delta H_{\rm eff}}}
\begin{equation}
\label{heffdef}
\tilde U_{\rm mult}(x)=U_{\rm mult}(x)\,e^{i\X(x)}\ . 
\end{equation}
From Eq.~(\ref{multproduct_static}) we may determine $\X(x)$ as:
\begin{equation}
\label{heffcalc}
e^{i\X(x)}=e^{i\delta H(n_m-1)}\cdot \ldots \cdot e^{i\delta H(1)}
\cdot e^{i\delta H}
\end{equation}
with 
\begin{equation}
\label{hjdeff}
\delta H(j)=U_{j-1}^{-1}\cdot\ldots\cdot U_1^{-1}\ \delta H
\ U_1\cdot\ldots\cdot U_{j-1}\ .
\end{equation}
We mention that the precise relation between $\X(x)$ and $\delta H$ is 
not really important in our approach since we directly model $\X(x)$ 
in our numerical simulations and use the expression 
(\ref{heffdef}) without entering into details of a particular 
implementation of $U_{\rm mult}(x)$. We remind that in (\ref{heffcalc}), 
(\ref{hjdeff}) the dependence of $\X(x)$ on $x$ is given by the choice 
of elementary gates $U_j$ which are changed with a change of $x$. 
A schematic quantum circuit of Shor's algorithm 
on a quantum computer with static imperfections in the quantum 
multiplication operator (\ref{eq:quantmultdef}) is shown in Fig.~\ref{fig2}.

\begin{figure}[h]
\epsfxsize=\hsize
\epsffile{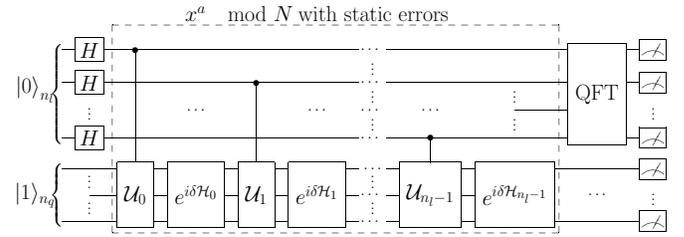}
\caption{\label{fig2}
Quantum circuit of Shor's algorithm on a quantum computer with static 
imperfections in the quantum multiplication operator 
${\cal U}_j=U_{\rm mult}\big(x^{2^j}\mod N\big)$ 
(\ref{eq:quantmultdef}) and the effective static perturbation 
$\delta {\cal H}_j=\X\big(x^{2^j}\mod N\big)$ 
[see Eqs.~(\ref{eq:quantmultdef}), (\ref{heffcalc})],
where in numerical simulations $\delta {\cal H}_j$ is
given by Eq.~(\ref{heffmodel}) with random realizations
of $\delta_i, J_i$ for practically each $j$ (see text). 
}
\end{figure}

As we already mentioned, the implementations of the quantum multiplication 
operator (\ref{eq:quantmultdef}) 
described in Refs.~\cite{paz1996,vedral} require additional work space 
qubits which are initially placed in the state $\ket{0\ldots 0}$ and 
are erased after the computation. The 
implementation of Ref.~\cite{paz1996} contains a quantum code that erases 
the work space qubits correctly but only for the relevant states 
$\ket{y}$ with $0\le y < N$ as initial states and not for 
the non-relevant states with $y\ge N$. For a perfect quantum computer this is 
of course not a problem, but when taking into account errors 
the non-relevant states 
may be populated and different implementations, which are absolutely 
equivalent for the relevant states, may potentially behave quite differently 
with errors. Even if the particular implementation ensures that a non-relevant 
state as initial state produces a properly erased work space register, 
the errors may still produce non-erased contributions. 

Actually the use of work space qubits implies that the notion of non-relevant 
states has to be enlarged, i.e. a combined state 
$\ket{y}\ket{\psi_{\rm work}}$ in the computational and work space 
register has to be considered as {\em non-relevant} if either 
$y\ge N$ for $\ket{\psi_{\rm work}}=\ket{0\ldots 0}$ or $y$ arbitrary for 
$\ket{\psi_{\rm work}}\ \perp\ \ket{0\ldots 0}$. If Shor's algorithm is 
implemented on a perfect quantum computer without any imperfections these 
non-relevant states are never populated. However, errors and imperfections 
will populate these states and their role is potentially quite important in 
this context. In this work we do not want to enter into the details 
of the effects due to the work space qubits. So, we simply assume that 
our model of imperfection effects (\ref{multproduct_static})  
acts only in the computational register, or in other words the 
static imperfections do not couple computational qubits with work space qubits.
However, even in this approximation we still keep track of the non-relevant 
states in the computational register (the states $\ket{y}$ with $y\ge N$).

For numerical simulations of 
Shor's algorithm in presence of imperfections
we use a classical computer taking into 
account the control register (with up to 20 qubits) and the computational 
register (with up to 10 qubits) and up to 30 qubits in total. 
We do not implement the quantum multiplication 
operator in terms of elementary gates but we directly implement the 
unitary operator as given in Eq.~(\ref{eq:quantmultdef}). 
To model the static imperfections we used the multiplication operator with 
errors given by (\ref{heffdef}) and with the effective perturbation operator
given by:
\begin{equation}
\label{heffmodel}
\X(x)=\sum_{i=0}^{n_q-1}\delta_i\sigma_i^{(z)}+2\sum_{i=0}^{n_q-2}J_i\sigma_i^{(x)}
\sigma_{i+1}^{(x)}
\end{equation}
where $\sigma_i^{(\nu)}$ are the Pauli operators acting on the $i$th qubit 
(of the computational register) 
and $\delta_j,\,J_j$ are random coefficients, chosen differently for each 
value of $x$ and distributed according to:
\begin{equation}
\label{Jdelta}
\delta_i,\ J_i\in[\sqrt{3}
\epsilon,\sqrt{3} \epsilon]\ .
\end{equation}
We remind that even for static imperfections $\X(x)$ 
given by Eqs.~(\ref{heffcalc}), (\ref{hjdeff}) strongly 
depends on the actual value of $x$ because this factor is hardcoded 
in realistic implementations by the choice of elementary gates $U_j$. 
According to Eq.~(\ref{eq:fullmult}), we have to apply the (controlled 
version) of the multiplication operator for all values 
\begin{equation}
\label{xvalues}
x\in\{x^{2^j}\mod N\ |\ j=0,\ldots,n_l-1\}\ .
\end{equation}
In our numerical simulations we have ensured by the proper choice of 
$\delta_i,\,J_i$ that $\X(x^{2^j} \mod N)$ 
is identical $\;\;$ to $\X(x^{2^l}\mod N)$ if for $j\neq l$ we 
have $x^{2^j}=x^{2^l}\mod N$, Otherwise, we have chosen different realizations
of $\delta_i, J_i$ for each value of 
$x^{2^j} \mod N$ assuming that the $x$-dependence of 
the hard coded implementation is sufficiently complex to render $\X(x)$ 
uncorrelated for different values of $x$. This introduces some kind of 
slight correlation that takes into account the static property of the 
imperfections. However, we have also checked that neglecting 
these correlations 
(choosing each time a different realization of $\X(x)$ even if 
the same $x$-value appears again) 
does not affect significantly our numerical results discussed below.
We also note that in potential applications (for ``real'' quantum computers) 
with larger values of $N$ and $n_l$, $n_q$ these kind of correlations 
will become less important. So, in majority of cases for each $j$
we have $\delta H_{\rm eff}(j)$ with independent random realizations
of $\delta_i, J_i$ in Eq.~(\ref{heffmodel}) distributed as in 
(\ref{Jdelta}).

\begin{figure}[htb!]
\epsfxsize=3.2in
\epsffile{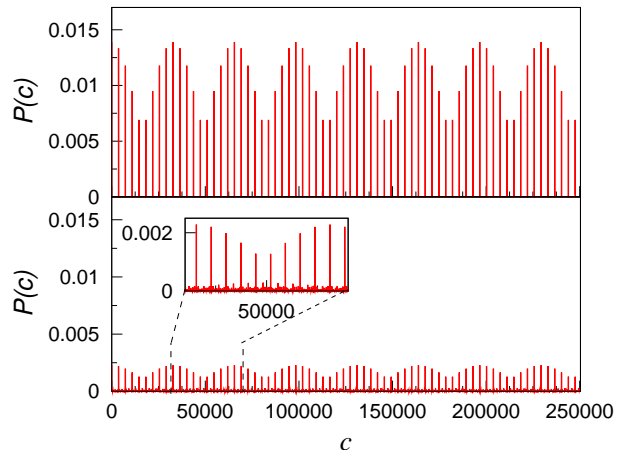}
\vglue -0.0cm
\caption{Probability 
$P(c)$ (Eqs.~(\ref{eq:prob},\ref{eq:probtot})) at the final stage
of Shor's algorithm for 
$\epsilon=0$ (top) and $\epsilon=0.1$ (bottom)
for values $N=323$, $n_q=9$, 
$L=27$, $x=2$, $r=72$.}
\label{fig3}
\end{figure}
In principle the unitary operator $e^{i\X(x)}$ is quite random due to 
Eqs.~(\ref{heffcalc}), (\ref{hjdeff}) and should directly couple many qubits 
in the computational register. Our model (\ref{heffmodel}) for the 
imperfections is quite convenient for numerical computations and is similar 
to the model used in \cite{benenti2001,frahm2004} 
but with a difference that in those works 
it is the elementary residual Hamiltonian $\delta H$ which is chosen in this 
way [see Eq.~(\ref{multproduct_static})]. Therefore we should expect 
that $\X(x)$ has a more complicated structure than (\ref{heffmodel}). 
However, choosing $\delta_j$ and $J_j$ of comparable size 
we are well in the quantum chaos regime 
\cite{georgeot2000,frahm2004,benenti2002} 
and therefore the model (\ref{heffmodel}) describes well 
the effects of static imperfections. 
It should be also noted that the quantum gates of the algorithm
introduce additional strong mixing between all qubits
even if they are not directly coupled by $\delta H_{\rm eff}(x)$
(see \cite{benenti2002} for details). We also checked
that a change of  $\delta H_{\rm eff}(x)$ from the form of Eq.~(\ref{heffmodel})
to the case when all qubits are coupled by residual interactions
does not affect significantly the results of numerical simulations.
This is in agreement with the results obtained in
\cite{frahm2004,benenti2002}.
Another advantage of a choice of $\delta H_{\rm eff}(x)$
in the form  (\ref{heffmodel}) is the local structure
of couplings between qubits that corresponds to 
a physical reality.
It is also important to note that when we have $n_g=n_l$ 
gates as in Fig.~\ref{fig2}, 
then the effective strength of $\epsilon$ is effectively renormalized
as $\epsilon \rightarrow \epsilon \sqrt{n_g}$
since static errors in each realization of  $\delta {\cal H}_j$
(see Fig.~\ref{fig2}) are independent and random.
We leave the question about possible strong correlations
between  $\delta {\cal H}_j$ due to 
a specific implementation
of the algorithm for future studies.

The above consideration assumes that sufficient randomization of static 
imperfections takes place along the path of a specific quantum circuit for 
the modular multiplication. In this case we may assume that the effective Hamiltonian 
$\delta H_{\rm eff}(x)$ in the propagator contains different random couplings between qubits
for each value of $x$ (see Eqs. (\ref{heffcalc}-\ref{Jdelta})). However, it is possible 
that the errors remain well correlated along the path of this circuit and in this 
case it is more appropriate to consider that $\delta H_{\rm eff}(x)$ does not depend on $x$ and 
remain the same along the whole Shor's algorithm. In our numeral studies we mainly 
concentrate on the first possibility (``generic imperfection model'') but in order to 
have the complete picture of the effects of static imperfections we also considered the 
second case with $\delta H_{\rm eff}(x)$ remaining constant along the full circuit 
(``correlated imperfection model''). According to our previous discussion 
the important property of both models is that the 
errors appear only via the positions of the propagator $e^{i \delta H_{\rm eff}(x)}$ between 
the modular multiplications in the full circuit of the algorithm. Hence the specific 
implementation of the modular multiplication circuit does not affect the random 
properties of inter-qubit couplings in $\delta H_{\rm eff}(x)$. Therefore, once the parametric 
dependence on the imperfection strength $\epsilon$, number of qubits $n_q$ 
and number of gates $n_g$ is established through numerical simulations, we can apply 
these results to arbitrary implementations currently discussed 
in the literature. 

In some sense, the model of static errors considered here can be 
viewed as a kind of generic static error model. It shows sufficiently 
rich and generic effects of errors and
due to its certain simplicity allows to make numerical
simulations with factorization of larger $N$ values
compared to previous numerical studies \cite{paz1996,paz1997,nori,hollenberg2}. 
This allowed us to determine the accuracy dependence on the parameters and to 
obtain the scaling law for a large number of qubits. This required to perform 
extensive numerical simulations with up to 30 qubits which became possible because we neglected 
the errors in the work space qubits. However, as soon as we obtain the parametric 
dependence of the algorithm accuracy we may reincorporate the effect of imperfections 
in the work space by modifying the effective qubit number in the computational 
register. We also neglected the static imperfections in the control register since 
the number of gates in the QFT (operating in the control register) 
is much smaller than the number of gates in the main part of Shor's algorithm. However, 
in the case of the correlated imperfection model, we verified that the introduction of 
couplings in the control register does not modify the established parametric dependence on 
the number qubits. We emphasize that our numerical calculations keep the exact quantum 
entanglement for the whole quantum evolution with up to 30 qubits. We note that a further 
increase of the factorized number $N$ can be achieved by replacing the control register 
by one qubit combined with appropriate measurements of this qubit and a semiclassical 
implementation of the QFT \cite{hollenberg2,mosca,zalka,plenio,beauregard}. However, 
this approach simulates the quantum measurement process in the algorithm and does not 
give a direct access to the full probability distribution in the quantum register which is 
substantially used in our studies. 

We present obtained numerical results 
in the next Section.
%

\section{Numerical results}
The effects of static imperfections in Shor's algorithm are studied
numerically following the approach described in the previous Section:
a wave vector of size $2^L$ is propagated numerically
according to the quantum circuits shown in Figs.~\ref{fig1},\ref{fig2},
all quantum gates are assumed to be exact, the imperfections,
induced by residual couplings between qubits 
in the computational register, are
encountered by the propagators $\exp (i \delta {\cal H}_j)$
appearing $n_l$  times in the circuit as it is described in Fig~\ref{fig2}.
We factorize numbers $N$ up to
$N\approx 1000$. This means that we simulate 
numerically a quantum computer with up to $30$ qubits, 10 computational
qubits and 20 control qubits 
(we assume ideal evolution in the workspace).
The list of factorized numbers $N$ used for numerical simulations
is given in Table I. We try to consider mainly  most
difficult cases when $N$ has only two factors 
and their values are more or less comparable.

\begin{table}[h]
\begin{ruledtabular}
\caption{Values for the data presented in Figs.~\ref{fig8},\ref{fig9}
 and \ref{fig10} for the generic imperfection model. 
Only the values with symbols are plotted in Figs.~\ref{fig8},\ref{fig9}. 
\label{table}}
\begin{center}
\begin{tabular}{crrrrrcc}
 $N$ & 	$n_q$ & $L$ & $\epsilon_c$ &$x$& $r$& &$\#$ real.\\
\hline 
$14$=$2\times 7  $& 4&12 & 	0.440& 3& 	6  &$\square$&45\\
$21$=$3\times 7  $&5 &15 & 	0.240& 2&	6&$\blacksquare$& 80\\
$33$=$ 3\times 11$&6 &18 & 	0.155 & 2& 	10& $\circ$&35\\
$35$=$ 5\times 7 $&6 &18 & 	0.157 & 4&	6& &60\\
$35$=$ 5\times 7 $&6 &18 & 	0.175 & 2&	12&$\bullet$&80 \\
$55$=$ 5\times 11$&6 &18 & 	0.155 & 6&	10& &70\\
$55$=$5\times 11$&6 &18 & 	0.175 & 2&	20& $\triangle$&80\\
$77$=$7\times 11$&7 &21 & 	0.155 & 10&	6& &70\\	
$77$=$7\times 11$&7 &21 & 	0.145 & 6&	10&$\blacktriangle$&70 \\
$77$=$7\times 11$&7 &21 & 	0.140& 	2&	30& &70\\
$91$=$7\times 13 $&7 &21 & 	0.135 & 3&	6& &70\\
$91$=$7\times 13 $&7 &21 & 	0.150& 	2&	12&$\triangledown$&35 \\
$143$=$11\times 13$& 8 &24 & 	0.115 & 2&	60& $\blacktriangledown$&35\\
$221$=$13\times 17$& 8 &24 & 	0.132 & 2&	24& $\lozenge$&50\\		
$299$=$13\times 23$& 9 &27 & 	0.106 & 2&	132& $\blacklozenge$&23\\		
$323$=$17\times 19$& 9 &27 & 	0.108 &2& 	72& +&30\\		
$437$=$19\times 23 $&	9 & 	27 & 0.099 & 2&	198&$\times$ &10\\		
$437$=$19\times 23$& 9 &27 & 	0.103 & 18&	22& &10\\
$505$=$5\times 101$&9 &27 & 	0.106 & 2&	100& $*$&10\\
$667$=$23\times 29 $&10 &30 &	0.098& 2&	308& $\square$&10\\
$943$=$23\times 41 $& 10&30 & 0.096&	2& 	220	&$\blacksquare$&10\\
\end{tabular}
\end{center}
\end{ruledtabular}
\end{table} 
In Fig.~\ref{fig3} we show a typical example 
of the probability distribution $P(c)$
of Eq.~(\ref{eq:probtot}) for the ideal case $\epsilon=0$ (top) and
for $\epsilon=0.1$ (bottom). It can be seen that 
the imperfections 
significantly reduce the amplitudes of the main $r$ peaks
and lead to appearance of new  
small peaks in new 
positions. 

\begin{figure}[htb!]
\epsfxsize=3.2in
\epsffile{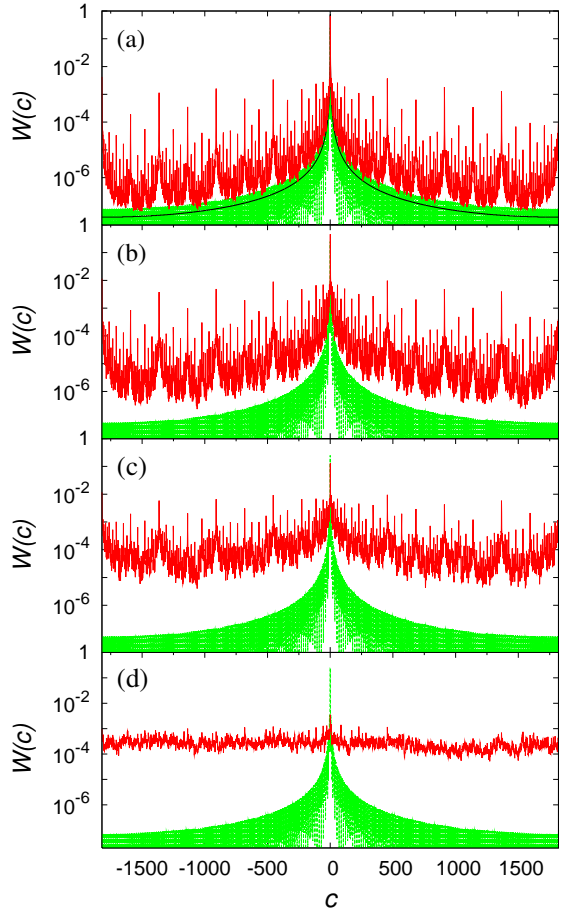}
\vglue -0.0cm
\caption{(color online) The global probability distribution $W(c)$,
as defined in Eq.~(\ref{eq:clash}), 
averaged over 10 realizations of random static imperfections, 
for different values of coupling strength $\epsilon$ : 
(a) $\epsilon=0.025$, (b) $\epsilon=0.05$,
(c) $\epsilon=0.1$, (d) $\epsilon=0.2$.  
The fast oscillating green (gray)
lower curve shows the theoretical probability
$G(c)$ at $\epsilon=0$ (Eq.~(\ref{eqgc}). 
The solid (black) curve in (a) is the actual
probability at $\epsilon=0$ obtained numerically.
The red (dark gray) curves show $W(c)$ obtained numerically at
given values of $\epsilon > 0$. Here, as in Fig.~\ref{fig3},
$N=323$, $n_q=9$, $L=27$, $x=2$ and $r=72$.}
\label{fig4}
\end{figure}
Since the success of the algorithm depends essentially 
on a probability of hitting $r -$~peaks in the process of measurement
then the most direct way  to study this probability is 
by {\em clashing\/} all the peaks into one, 
or in other words, adding them 
all together by taking $c$ modulus $s$ where $s$ is the nearest
integer value of the ratio $Q/r$ and thus reducing all probabilities
inside one cell with $s$ states. 
In this way we obtain a new distribution of global search probability  $W(c)$:
\begin{equation}        
        \label{eq:clash}
W(c)=\sum_{j=0}^{r-1} P([c+s+jQ/r] \mod s)
\end{equation} 
where now $c=-s/2,\ldots,s/2-1$ (the difference of 
$c$ for $P$ and $W$ is clear from the context)
and $s \approx Q/r$ is the distance between peaks.
For the ideal algorithm this global probability
$W(c)$ has one peak at $c=0$ that stresses the important
property of Shor's algorithm: it is not important 
what peak from the main chain of $r$ peaks
is selected by measurement, it is important
to know its exact position modulus $s$
that allows to determine $r$ value and then
to find the factors of $N$ by classical
computations. The global probability $W(c)$
is distributed over states with $c=-s/2,\ldots,s/2-1$
and is normalized to unity in this interval.

In Fig.~\ref{fig4}  we show a typical example
of the global probability $W(c)$ variation with 
the increase of coupling strength $\epsilon$.
The distribution $W(c)$ for the ideal algorithm at $\epsilon=0$
is well described by the envelope 
function $W_0(c) = (\sin(\pi c)/(\pi c))^2$
of the distribution $G(c)$ discussed in  \cite{gerjuoy}
(see also (\ref{eq:probresult}) ):
\begin{equation}
\label{eqgc}
G(c)=\left(\frac{r}{Q}\right)^2\left(\frac{\sin(\pi c)}
{\sin(\pi c r/Q)}\right)^2 \; .
\end{equation}

\begin{figure}[htb!]
\epsfxsize=3.2in
\epsffile{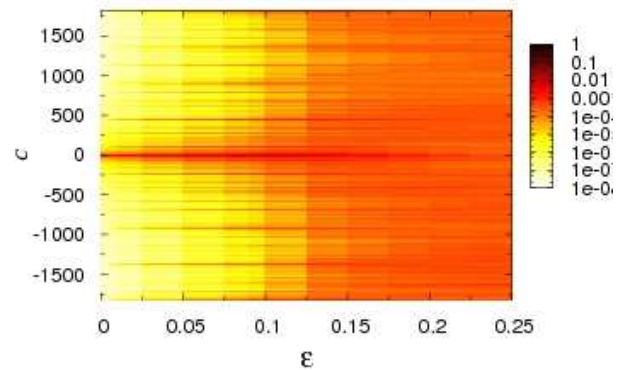}
\vglue -0.0cm
\caption{(color online) Quantum melting of Shor's algorithm
induced by imperfections: color density plot of the 
global search probability $W(c)$ as a function of
coupling strength between qubits $\epsilon$
for $N=323$, $n_q=9$, $L=27$, $x=2$, $r=72$ 
($W(c)$ is averaged over 20 realizations).}
\label{fig5}
\end{figure}

Shor's algorithm is successful if the
probability at $c = 0$ is significant (comparable
to 1). This is indeed the case for small
values of $\epsilon$ (Fig.~\ref{fig4}a,b).
In these cases the main probability is concentrated
near $c=0$. There are new peaks appearing at 
very large values of $c$ but they have rather small total
probability. With a further growth of $\epsilon$
the number of such peaks and their probability
grow (Fig.~\ref{fig4}c), the amplitude of the central peak at $c=0$
drops and above certain $\epsilon$ the distribution
$W(c)$ becomes practically flat (Fig.~\ref{fig4}d)
that signifies the complete destruction of the algorithm.
A pictorial view of variation of $W(c)$ with $\epsilon$
is shown in Fig.~\ref{fig5}.

\begin{figure}[htb!]
\epsfxsize=3.2in
\epsffile{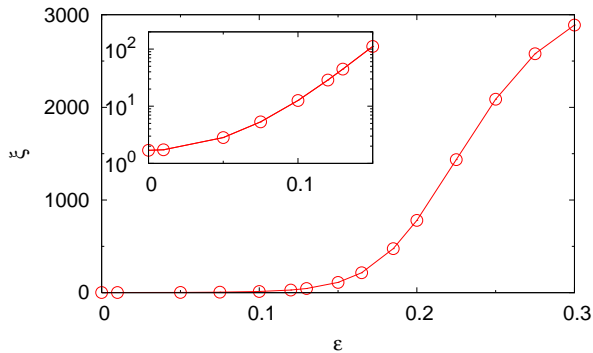}
\vglue -0.0cm
\caption{Averaged IPR $\xi$ given by Eq.~(\ref{eqipr}) 
as a function of $\epsilon$
for $N=323$, $n_q=9$, $L=27$, $x=2$, the inset shows 
the dependence on small $\epsilon$ in log-scale,
average is done over the number of realizations given in Table I.}
\label{fig6}
\end{figure}

\begin{figure}[htb!]
\epsfxsize=3.2in
\epsffile{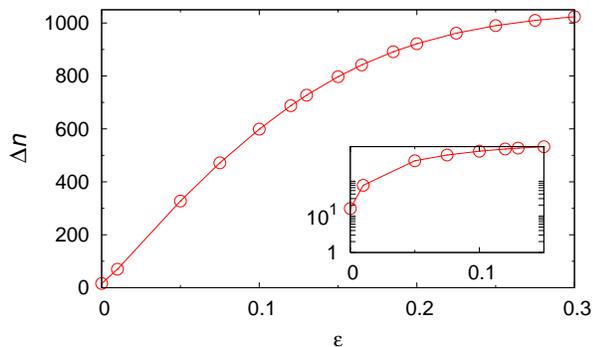}
\vglue -0.0cm
\caption{Same as in Fig.~\ref{fig6} but for $\Delta n$ given by
Eq.~(\ref{eqdeltan}).}
\label{fig7}
\end{figure}

In order to study the effects of static imperfections on 
the algorithm accuracy in a more quantitative way 
it is convenient to use
the inverse participation ratio (IPR) 
\begin{equation}
\label{eqipr}
\xi=\sum_c |W(c)|^{-2}
\end{equation}
which gives a number of effectively
populated states in the distribution $W(c)$.
This quantity is extensively used to characterize the properties
of many-body quantum states 
(see e.g. \cite{georgeot2000,georgeot1997}).
Another convenient characteristics is the width of the distribution
defined as
\begin{equation}
\label{eqdeltan}
\Delta n=\sqrt{\sum_c W(c)(c-\langle c\rangle)^2} \; .
\end{equation}
The dependence of these quantities on the perturbation strength $\epsilon$
is shown in Figs.~\ref{fig6} and \ref{fig7} for the typical case $N=323$.
The value of $\xi$ is practically constant up to a value 
$\epsilon_c \approx 0.1$ after which it starts to
grow abruptly. On a contrary, the width $\Delta n$ 
grows starting from small values of $\epsilon$. At large $\epsilon$
the saturation of growth takes place due to a finite
number of states inside  the distribution $W(c)$.

\begin{figure}[ht!]
\epsfxsize=3.2in
\epsffile{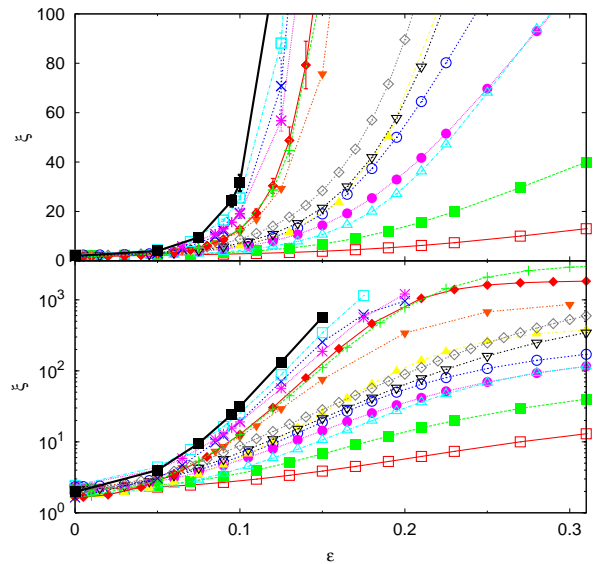}
\vglue -0.0cm
\caption{(color online) Dependence of averaged IPR $\xi$ 
on strength of imperfections $\epsilon$
for different values of $N$, curves with symbols
from the list of Table I (from top curves with the largest $N=943$
to bottom curves with the smallest $N=14$);
number of disorder realizations used for averaging is
given in the Table. For the large values of $N$
we show typical statistical error bars,
for small $N$ the error bars are comparable with the symbol size
and we do not show them. Top and bottom panels show
$\xi$ in normal and logarithmic scale respectively.}
\label{fig8}
\end{figure}

\begin{figure}[ht!]
\epsfxsize=3.2in
\epsffile{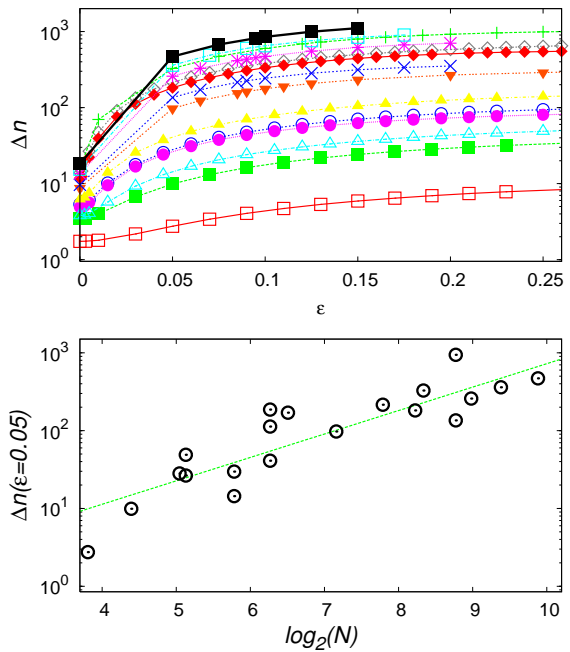}
\vglue -0.0cm
\caption{(color online) Top panel: dependence of averaged 
$\Delta n$ on $\epsilon$
for different $N$ with the same symbols from Table I as in Fig.~\ref{fig8}.
Bottom panel shows the dependence of $\Delta n$ on $N$
in log-log scale
for $\epsilon = 0.05$, the straight line shows
the dependence $\Delta n = A \epsilon N$ with $A \approx 14$.
}
\label{fig9}
\end{figure}

The dependence of the IPR $\xi$ on $\epsilon$
for different $N$ is shown in Fig.~\ref{fig8}.
The data clearly show that the dependence 
becomes more and more sharp with the increase of $N$.
For $\Delta n$ we see a strong increase with $N$
but there is no such sharp behavior (see Fig.~\ref{fig9}).
We attribute such a difference to the fact that 
even small $\epsilon$ gives far transitions with
exponentially large $c \sim s \approx Q/r \propto 2^{n_q}$.
Due to that the second moment of the probability
distribution  grows exponentially with the number of qubits.
The numerical data on dependence of $\Delta n$ on $N$
at a small fixed $\epsilon$ indeed show the
exponential growth with $\Delta n \approx A \epsilon  N$
with a numerical constant $A \approx 14$ 
(Fig.~\ref{fig9} bottom panel).  
A similar behavior has been seen for quantum chaos
algorithms \cite{song,levi}. 
The mechanism of this exponential growth is the following
\cite{song,levi}: the gates with imperfections
transfer a probability $W_\epsilon \sim \epsilon^2 n_q$
from the search state at $c \approx 0$
to about $n_q$ peaks (see Fig.~\ref{fig4})
distributed in the interval of size $s \sim N$.
Here, $n_q$ comes from the norm of the
Hamiltonian (\ref{heffmodel}) with $n_q$ qubits
with local couplings.
There are $n_g = n_l \approx 2 n_q$ such transitions $W_\epsilon$
during the whole algorithm computation.
Thus, we obtain the second moment of the distribution $W(c)$:
\begin{equation}
\label{eqexp}
(\Delta n)^2 \approx a^2 \epsilon^2 n_q N^2 \; ,
\end{equation}
where according to numerical data of Fig.~\ref{fig9}
(bottom) the numerical coefficient $a \approx A/\sqrt{n_q} \approx 4.5$
is close to the one obtained in \cite{song,levi}.
Of course, the fluctuations in Fig.~\ref{fig9} (bottom)
are rather large. We think that the main origin
of these fluctuations is related to the arithmetic
properties of $x, r$ and $N$. Indeed, $r$ varies significantly
with $x$ and $N$ (see Table I) that clearly affects the
transition probability induced by imperfections \cite{note}.
In spite of these fluctuations the global exponential growth
of $\Delta n$ with $n_q$ is seen rather clearly.
Such an exponential sensitivity of $(\Delta n)^2$
on $N$ is not very pleasant for the algorithm accuracy,
but in principle this behavior is not so dangerous.
Indeed, the total probability to have 
exponentially large values of $c$ is very small
and doing a few measurements and making a majority ``vote''
will eliminate such extreme values of $\Delta n$.

Therefore, more crucial is the behavior of $\xi$
since above a certain critical value $\epsilon_c$
the probability $\; W(c) \;$ spreads over very many levels
and the algorithm stops to work.
Indeed, it is known that static imperfections
can lead to a complete delocalization,
for example, in the case of a quantum algorithm
simulating the Anderson localization in
three dimensions \cite{pomeranskyand}.

\begin{figure}[ht!]
\epsfxsize=3.2in
\epsffile{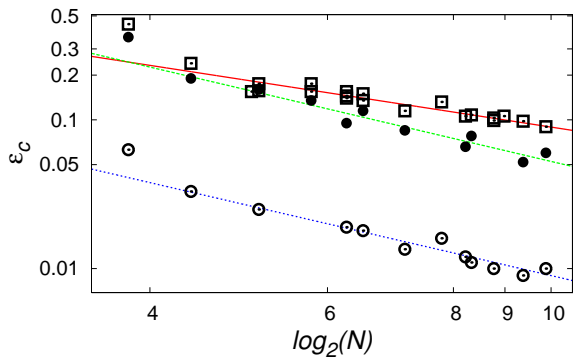}
\vglue -0.0cm
\caption{Dependence of $\epsilon_c$, obtained from  the criterion
$\xi(\epsilon_c) = 10 \xi(\epsilon=0)$, 
on $\log_2 N$ in log-log scale. 
The numerical data are shown for 
the generic imperfection model (squares, data from Table I), 
the correlated imperfection model 
with qubit couplings in the computational register (full circles) and 
the correlated imperfection model 
with all qubits in the control and computational registers coupled by 
interactions (open circles), see text for model description.
The straight lines show the fit $\epsilon_c = B/(\log_2 N)^{\beta}$
in the interval $4 < \log_2 N <10$ with
$B=0.98 \pm 0.16$, $\beta =1.04 \pm 0.094$ for squares (top line), 
$B=2.06 \pm 0.42$, $\beta =1.6 \pm 0.11$ for full circles (middle line) 
and 
$B=0.33 \pm 0.05$, $\beta =1.57 \pm 0.09$ for open circles (bottom line). 
}
\label{fig10}
\end{figure}

To determine the delocalization border 
for Shor's algorithm and the dependence of $\epsilon_c$
on $N$ in the generic imperfection model we use a numerical criterion
$\xi(\epsilon_c)=10\xi(\epsilon=0)$.
Indeed, an increase by a factor 10 is sufficiently large
to obtain the transition border in $\epsilon$.
The dependence of $\epsilon_c$ on $N$ is shown in
Fig.~\ref{fig10} (squares and top line). From the theoretical view point
the errors are accumulated randomly so that the
probability $W_t$ transferred from $c = 0$ 
to all other states grows proportionally to the number of gates 
$n_g$ with errors and thus  
$W_t \sim W_{\epsilon} n_g \sim \epsilon^2 n_q n_g
\sim \epsilon^2 n_q^2$.
We expect that above the border $W_t \sim 1$ 
the probability becomes delocalized  over exponentially many
states and the algorithm is destroyed.
This gives the quantum chaos border
\begin{equation}
\label{eqepsc}
\epsilon_c(N)= B/\log_2 (N) \approx \sqrt{2} B/\sqrt{n_q n_g}  \; ,
\end{equation}
where $B$ is a numerical constant. For our generic imperfection
model we have in the second equality $n_g \approx 2 n_q$ 
but in the case when the errors related with the workspace
qubits are taken into account we have 
$n_g \sim n_q^3 \sim ({\log_2 N})^3$. 
The numerical data  for $\epsilon_c$ are presented in Fig.~\ref{fig10}.
The fit of the dependence in the form
$\epsilon_c = B/(\log_2 N)^{\beta}$ in the interval $4 < \log_2 N <10$
gives $B=0.98 \pm 0.16 $, $\beta =1.04 \pm 0.094 $.
Thus, the numerical data confirm the theoretical
estimate (\ref{eqepsc}) with $B \approx 1$.
The border $\epsilon_c$ drops polynomially with $\log_2 N$
since the whole Shor algorithm is performed in a polynomial number of gates
$n_g \sim (\log_2 N)^3$.
In this respect the situation is different from the case
of the Grover algorithm with imperfections considered in \cite{zhirov}
where the number of gates grows exponentially with $n_q$.
The exponential sensitivity of Floquet eigenstates to
static imperfections in quantum chaos algorithms \cite{benenti2002} 
also corresponds to a different situation
since in a sense eigenstate corresponds to a very long time scale
where the number of gates becomes exponentially large.

In the above consideration for the generic imperfection model, we 
assumed that the quantum circuit effectively modifies the couplings 
between qubits in the propagator from one gate to another. Another 
limiting case corresponds to the correlated imperfection model, 
where these couplings remain unchanged from gate to gate (see Section 
\ref{sec3}). For this particular model we also performed extensive numerical 
simulations considering two cases a) the interactions exist only 
between qubits in the computational register (see Fig. \ref{fig10}, 
full circles, middle line) and b) the interactions exist between 
all qubits in the control and computation registers (see Fig. \ref{fig10}, 
open circles, bottom line). For the numerical study of these two cases we 
used the same quantities as those described above for the generic imperfection 
model. We do not reproduce all data here but only show the cumulative 
final dependence for the quantum chaos border $\epsilon_c$ defined by the 
same relation $\xi(\epsilon_c) = 10 \xi(\epsilon=0)$.
The fit of the numerical data in the form
$\epsilon_c = B/(\log_2 N)^{\beta}$ gives the same exponent $\beta\approx 1.6$ 
for both cases of the correlated imperfection model with the numerical 
factors as in Fig. \ref{fig10}. Naturally $B$ becomes smaller when all 
qubits are coupled. The value of $\beta$ is definitely larger 
as compared to the 
generic imperfection model (where $\beta\approx 1$). This can be understood 
on the following physical grounds: the errors accumulate coherently along 
$n_g$ gates so that the transition probability from the target state to 
all other states is $W_t \sim W_{\epsilon} n_g^2 \sim \epsilon^2 n_q n_g^2
\sim \epsilon^2 n_q^3$. The quantum chaos border is given by the condition 
$W_t \sim 1$ that gives:
\begin{equation}
\label{eqepsc2}
\epsilon_c(N)= B/\log_2 (N)^{3/2} \approx 2 B/\sqrt{n_q n_g^2}  \; ,
\end{equation}
since we always chose $n_g \approx 2 n_q$. The theoretical exponent 
$\beta=1.5$ is in good agreement with the numerical fit $\beta=1.6 \pm 0.1$.
We also clearly see that the fact of coupling all qubits does not affect 
the parametric dependence of the chaos border on $\log_2 (N)$ and gives only a 
change of the numerical prefactor $B$. It is important to note that the 
quantum chaos border is lower for the correlated imperfection model.
\section{Conclusion}
We performed extensive numerical simulations of Shor's algorithm
factorizing numbers up to $N=943$ on a quantum computer 
with up to 30 qubits in presence of residual static couplings
between qubits. Our studies show that the width $\Delta n$
of $r$-peaks, which positions are essential for determination
of factors of $N$, grow exponentially with $N$
(see Eq.~(\ref{eqexp})).
However, the use of majority vote with few measurements
allows to eliminate the rare events which contribute to this
exponential growth. In fact the algorithm remains
operational up to the critical coupling strength $\epsilon_c$
which drops polynomially with $\log_2 N$ (see Eq.~(\ref{eqepsc})).
Since with the work space qubits the total
number of gates in Shor's algorithm is
$n_g \sim (\log_2 N)^3$ the relation (\ref{eqepsc})
gives $\epsilon_c \sim 1/(\log_2 N)^2$. 
In this estimate,
based on Eq.(\ref{eqepsc})
with $n_q \sim \log_2 N$ and $n_g \sim (\log_2 N)^3$,
we assume the validity of the generic imperfection model
where couplings fluctuate from gate to gate.
Another limit corresponds to the case of correlated imperfection model
where couplings remain fixed for all gates.
In this case the relation (\ref{eqepsc2})
gives $\epsilon_c \sim 1/(\log_2 N)^{7/2}$.
A presence of finite correlation length  $1 \leq n_{gcor} \leq n_g $ in the number of gates
$n_g$ will give interpolation between these two limiting cases
with $\epsilon_c \sim 1/[(\log_2 N)^2 \sqrt{n_{gcor}}]$.
At present,
the latest RSA challenge number factored is 
RSA-640 with $\log_2 N = 640$ \cite{rsa}. Thus, 
assuming a more optimistic case of the generic imperfection model, a quantum
computer which factors this number should
have a dimensionless 
coupling strength $\epsilon < \epsilon_c \sim 2 \times 10^{-6}$.
The  value of $\epsilon$ can be interpreted as
$\epsilon \approx J_{\rm res} \delta t$, where $J_{\rm res}$ is a strength
of residual couplings and $\delta t \approx 1/J_{g}$ is a time duration
of two-qubit gate which is related to a typical value
of coupling $J_{g}$ between two qubits which
implements this gate. As a result, we obtain
that $\epsilon \sim J_{\rm res}/J_{g}$ has the meaning
of the ratio between a residual coupling between
qubits and a coupling strength implementing a two-qubit gate.
According to the above estimate in a quantum computer this
ratio should be kept as small as 
$J_{\rm res}/J_g < \epsilon_c \sim 2 \times 10^{-6}$
to have a possibility to beat a modern classical computer
in the RSA-factorization. Such a restriction rises
serious requirements to experimental implementations of
quantum computers, but it's possible to hope that
future technological progress will make this possible.
Finally we note that we do not consider here
quantum error corrections (see \cite{gottesman} and Refs. therein)
which may improve the situation but on a price of significant
increase of the total number of qubits required for
computations.

This work was supported in part by the EC IST-FET project EuroSQIP.
For numerical simulations we used the codes of Quantware Library \cite{qwlib}.

\appendix
\section{Erratum}
\begin{figure}[h!]
\begin{center}
\includegraphics[width=8cm]{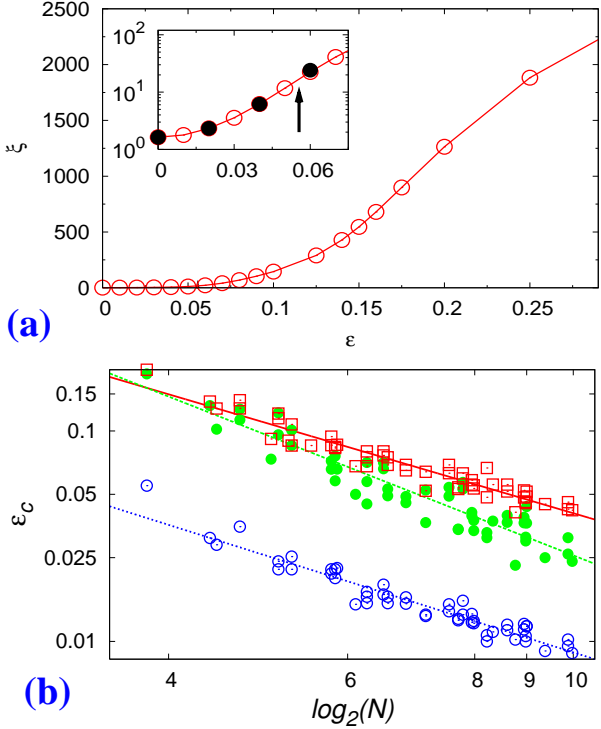}
\end{center}
\caption{(Color online)
{\bf \blue (a)} The open circles ($\circ$) represent the averaged IPR $\xi$ 
as a function of $\epsilon$
for $N=323$, $n_q=9$, $L=27$, $x=2$, the inset shows 
the dependence for small $\epsilon$ in log-scale, the
average is done over twenty disorder realizations. The full circles
($\bullet$) show data obtained by a different numerical method [I. Garc\'\i a-Mata, K.M. Frahm, and
D. L. Shepelyansky, in preparation]. The black arrow indicates $\epsilon_c$ defined by 
the criterion $\xi(\epsilon_c) = 10 \xi(\epsilon=0)$.
{\bf \blue (b)}
Dependence of $\epsilon_c$ on $\log_2 N$ in a logarithmic scale. 
The numerical data points correspond to
the generic imperfection model (open squares), the correlation imperfection model (full circles), 
and the the correlated imperfection model with all qubits coupled (open circles).
The statistical error due to different disorder realizations is smaller than the symbol size. 
The straight lines show the fit $\epsilon_c = B/(\log_2 N)^{\beta}$ (see text). 
\label{fig:ERR}}
\end{figure}

In the published version [Phys. Rev. A {\bf 75}, 052311 (2007)] as well as 
in the original arXive version there is an error 
in the numerical implementation of the static imperfection model (Eq.~(21) in the
paper) which resulted in an effective reduction of the disorder strength 
$\epsilon$ by approximately a factor of $2$ for the cases of generic and correlated imperfection 
models.
However, there was no error for the correlated 
imperfection model with all qubits in the control and computational register coupled
by interactions. 
After correction of the error the dependence of the inverse participation ratio
(IPR) $\xi$ on $\epsilon$ (see Fig.~\ref{fig:ERR} {\bf \blue (a)}) remains 
qualitatively the same as in Fig.~6 of the paper. 
Nevertheless, the values of $\epsilon_c$ fluctuate strongly depending on the arithmetic 
properties of $x, r$ and $N$. These fluctuations remain quite strong even after increasing the 
number of data points (see Table II and Fig.~\ref{fig:ERR} {\bf \blue (b)} as compared to 
Table I and Fig.~10). 
The algebraic fit $\epsilon_c = B/(\log_2 N)^{\beta}$ gives 
$\ln B=0.068 \pm 0.105$, $\beta =1.420 \pm 0.054$ for the generic imperfection model, 
$\ln B =0.70 \pm 0.19$, $\beta =1.897 \pm 0.097$ for the correlated imperfection model, 
and 
$\ln B =-1.22 \pm 0.13$, $\beta =1.523 \pm 0.068$ for the correlated imperfection model with all 
qubits coupled. 
We attribute the deviations of the exponents $\beta$ from the theoretical values to finite size 
effects in number of qubits and the significant arithmetic fluctuations of $\epsilon_c$ as a function 
of $r$, $x$ and $N$. 

In addition after publication we realized there were two typos in Eqs. (22) and (26) which 
should read
\begin{displaymath}
\label{Jdelta}
\delta_i,\ J_i\in[-\sqrt{3}
\epsilon,\sqrt{3} \epsilon]\quad;\quad
\xi=\left(\sum_c |W(c)|^{2}\right)^{-1}\ .
\end{displaymath}
This did not affect any of the calculations.

\begin{table}[h!]
\caption{Values for the data presented in Fig.~\ref{fig:ERR} {\bf \blue (b)} with 
$\epsilon_c^{(1)}$ for the generic imperfection model ({\red$\square$}),
$\epsilon_c^{(2)}$ for the correlated imperfection model({\green$\bullet$}) and 
$\epsilon_c^{(3)}$ for the correlated imperfection model with all qubits coupled 
({\blue $\circ$}).
$N_R$ is the approximate number of random realizations for each case. 
\label{table2}}
{\small
\begin{ruledtabular}
\begin{tabular}{|l|c|c|c|c|c|c|c|c|c|}
$N$&$n_q$&$L$& $\epsilon_c^{(1)}$& $\epsilon_c^{(2)}$&$\epsilon_c^{(3)}$& x &r & $N_R$\\ \hline
$14=2\times 7$&4&12& 0.1955 &0.187&0.055& 3 &6 &40 \\ 
$21=3\times 7$ &5&15& 0.1380 &0.132&0.031& 2 &6 &40\\ 
$22=2\times 11$ &5&15& 0.1279 &0.102&0.0288& 7 &10 &40 \\ 
$26=2\times 13$ &5 & 15& 0.140 &0.127&0.0351& 11 &12 &40 \\ 
$26=2\times 13$&5 &15 & 0.128 &0.113&0.024& 17 &6&40 \\
$33=3\times 11$ &6 & 18& 0.0917&0.0735&0.0175& 2  &10&40 \\
$35=5\times 7$ &6 &18 & 0.121 &0.122&0.024& 2 &12&40\\ 
$35=5\times 7$&6 & 18& 0.115 &0.096&0.022& 4 & 6 &40\\ 
$39=3\times 13$&6 &18 & 0.1074 &0.101&0.0253& 2 &12 & 40 \\ 
$39=3\times 13$&6 &18 & 0.0853 &0.0853&0.022& 4 &6&40\\ 
$55=5\times 11$ &6 &18 & 0.0940 &0.072&0.022& 2 &20&30\\ 
$55=5\times 11$&6 &18 & 0.0856 &0.0658&0.021& 4 &10&30\\ 
$57=3\times 19$ &6 & 18& 0.083 &0.058&0.02& 2 & 18 &20\\ 
$57=3\times 19$&6 &18 & 0.0084 &0.076&0.022& 8 &6&20\\
$58=2\times 29$ &6 &18 & 0.086 & 0.0659 &0.0223& 3 &  28&20\\ 
$69=3\times 23$ &7 &21 & 0.068 &0.050&0.015& 2 &22&40\\ 
$77=7\times 11$ &7 &21 & 0.068 &0.045&0.0152& 2 &30&40\\ 
$77=7\times 11$&7 &21 & 0.075 &0.0575&0.0163& 6 &10&40\\ 
$77=7\times 11$&7 &21 & 0.080 &0.0708&0.0172& 10 &6&40\\ 
$91=7\times 13$ &7 &21 & 0.080 &0.0707&0.0186& 2 &12&30\\ 
$91=7\times 13$ &7 &21 & 0.0765 &0.0661&0.0167& 3 &6&30\\ 
$95=5\times 19$ &7 &21 & 0.075 &0.053&0.0163& 2 &36&40\\ 
$95=5\times 19$&7 &21 & 0.069 &0.0492&0.0152& 4 &18&40\\ 
$115=5\times 23$ &7 & 21& 0.069  &0.0488&0.0162& 2 & 44&16\\ 
$115=5\times 23$&7& 21& 0.065&0.0456&0.0151& 4 &22&16\\ 
$143=11\times 13$ &8 & 24& 0.052&0.0367&0.0132& 2 &60 &30\\ 
$143=11\times 13$&8 &24 & 0.064&0.0524&0.0134& 8 & 20&30\\
$187=11\times 17$ &8 &24 & 0.069&0.054&0.015& 2 &40&20\\ 
$187=11\times 17$&8 &24 & 0.063&0.049&0.0139& 4 &20&20\\ 
$209=11\times 19$ &8 &24 & 0.054&0.034&0.0126& 2 &90 &20\\ 
$209=11\times 19$ &8 &24 & 0.053&0.0342&0.0128& 7 &30&20\\ 
$221=13\times 17$ &8 & 24& 0.0629&0.057&0.0156& 2 &24&20\\ 
$221=13\times 17$&8 &24 & 0.0595&0.053&0.0134& 4 & 12&20\\ 
$247=13\times 19$&8 &24 & 0.0584&0.041&0.0135& 2 &36 &20\\ 
$247=13\times 19$&8 &24 & 0.0556&0.0395&0.0123& 3 &18& 20\\ 
$253=11\times 23$ &8 &24 & 0.0533&0.0336&0.0122& 2 & 110& 20\\ 
$253=11\times 23$&8 &24 & 0.0552&0.0376&0.0125& 10 &22& 20 \\ 
$299=13\times 23$ &9 &27 & 0.0667&0.0329&0.0107& 2 &132&20\\ 
$299=13\times 23$&9 &27 & 0.0484&0.0311&0.010& 4 &66& 20\\ 
$323=17\times 19$ &9 &27 & 0.0556&0.0373&0.0111& 2 &72& 20\\ 
$391=17\times 23$ &9 &27 & 0.0516&0.0368&0.0113& 2 &88&20 \\ 
$391=17\times 23$&9 &27 & 0.0556&0.0397&0.0120& 3 &176& 20\\ 
$437=19\times 23$ &9 &27 & 0.0411&0.023&0.0102& 2 &198 & 16\\ 
$493=17\times 29$ &9 &27 & 0.0483& 0.0366 &0.0114& 2 &56 & 16\\ 
$493=17\times 29$&9 &27 & 0.0522& 0.0389 &0.0121& 3 &112&16\\ 
$505=5\times 101$&9 &27 & 0.046 &0.0313&0.0106& 2 &100& 16\\ 
$505=5\times 101$&9 &27 & 0.0454&0.030 &0.010& 4 &50& 16\\ 
$511=7\times 73$&9 &27 & 0.0488&0.0437&0.0118  & 3 &12 & 16\\ 
$511=7\times 73$&9 &27 & 0.0515&0.0365&0.0118  & 5 &72& 16\\ 
$667=23\times 29$ &10 &30 & 0.0449&0.025&0.0090& 2 &308 & 10\\ 
$943=23\times 41$ &10 &30 & 0.0425&0.026&0.0094& 2 &220& 10\\ 
$1007=19\times 53$ &10 &30 & 0.042  &0.024&0.0088& 2 &468& 10
\end{tabular}
\end{ruledtabular}
}
\end{table}

\end{document}